\documentclass[pre,aps,amsfonts,amssymb,amsmath,showpacs,twocolumn]{revtex4}

\usepackage{times} 

\usepackage{graphicx}
\usepackage{bm}
\usepackage{color}
\usepackage{amsfonts}
\usepackage{amssymb}
\usepackage{amsthm}

\definecolor{pink}{rgb}{1,1,0} 
\definecolor{red}{rgb}{1,0,0}
\definecolor{yellow}{rgb}{1,1,0}
\definecolor{orange}{rgb}{1,0.5,0}
\definecolor{white}{rgb}{1,1,1}
\definecolor{blue}{rgb}{0,0,1}

\begin{document}

\title{ Master Equation approach to Reversible and Conservative Discrete Systems }

\author{ Felipe Urbina and Sergio Rica}

\address{ Facultad de Ingenier\'ia y Ciencias, Universidad Adolfo Ib\'a\~nez, Avda. Diagonal las Torres 2640, Pe\~nalol\'en, Santiago, Chile.}

\begin{abstract}
A master equation approach is applied to a reversible and conservative cellular automata model (Q2R). The Q2R model is a dynamical variation of the Ising model for ferromagnetism that possesses quite a rich and complex dynamics. The configurational space is composed by a huge number of cycles with exponentially long periods. Following Nicolis and Nicolis [\textit{Phys. Rev.} \textbf{A 38}, 427-433 (1988)], a coarse-graining approach is applied to the time series of the total magnetization leading to a master equation that governs the macroscopic irreversible dynamics of the Q2R automata. The methodology is replicated for various lattice sizes. In the case of small systems, it is shown that the master equation leads to a tractable probability transfer matrix of moderate size which provides a master equation for a coarse-grained probability distribution. The method is validated and some explicit examples are discussed. 
\end{abstract}

\date{ \today}

\maketitle

\section{Introduction}

\noindent

In statistical physics one basically considers a large set of reversible and conservative ordinary differential equations for the description of particle dynamics. The temporal evolution for this cumbersome problem, even for a modest number of particles, requires a statistical description which introduces the concept of probability distribution function (PDF) for the phase space of the system. Irreversibility, equilibrium, and more important non-equilibrium properties, surge from this probability conception of systems (with a  large number of degrees of freedom) and its deterministic evolution. Briefly, the methodology reduces (under some assumptions) to a kinetic description which displays an irreversible behavior to equilibrium observed in macroscopic systems.
The assumptions for this approach are:  i) Macroscopically, a system is described by a finite set of observables; ii) The robust instability of the microscopic motions which is at the basis of the sensibility to initial conditions and the ergodic assumption.
iii) A {\it Stosszahlansatz} that introduces explicitly a broken before-after symmetry of the probability distribution evolution.

About 20 years ago, Nicolis {\it et al.} \cite{nicolis,et} introduced a systematic coarse-graining approach to the treatment of the macroscopical variables.  As a consequence, this coarse-graining breaks naturally the past-future symmetry in time, leading to an irreversible master equation, for a reduced probability distribution function of the system. 
In the present paper, we apply this systematic approach to a conservative and explicit reversible cellular automata.
In particular, we  consider the  Q2R  model, introduced by Vichniac in the mid 80s \cite{vichniac}, which is a cellular automata that runs on a two dimensional grid of finite size, and is reversible in a physical sense, that is, the automata rule is not only invertible, but the backward rule reads exactly the same as the forward one. Moreover, it was shown by Pomeau \cite{pomeau84} that the Q2R automata  possesses a conserved energy-like quantity. 

The main reason to apply  the coarse-graining approach to a cellular automata instead to a coupled system of ordinary differential equations, is because a cellular automata is a discrete model with boolean entities as microscopic variables, thus, the system is numerically reversible and conservative. In consequence, Q2R seems to be a good benchmark to test the principles of statistical physics.  
But, the phase space is finite, hence the dynamical system only possesses fixed points and periodic orbits, therefore it cannot be ergodic, at least in the usual sense of continuos dynamics. However, for large enough  systems, the phase space becomes huge, and the periodic orbits may be, as we show, exponentially long, so in practice, of infinite period. Further, if the initial state is random, the temporal behavior may be quite random and it possesses many aspects of chaotic systems, as sensibility to initial conditions, mixing, etc. For any purpose, the observation of a short periodic orbit is really improbable for large enough systems with random initial conditions. In general, there is a huge number of initial conditions that  are ``almost'' ergodic. Numerical studies show that the premises of statistical physics are valid, in particular the observables may be computed using the standard methods of statistical physics. 

The study of the dynamics and properties of the Q2R model has a long history.  Soon after the seminal works of Vichniac \cite{vichniac} and Pomeau \cite{pomeau84}, Herrmann\cite{hans} implemented the Q2R algorithm to study the two space dimensional Ising model in the frame of the micro-canonical ensemble. He studied the global magnetization, obtaining an excellent  representation for the magnetization as a function of the initial conserved energy, displaying a coherent picture for the phase transition of the Ising model. Later, Herrmann, Carmesin and Stauffer \cite{hans2} studied numerically the probability to reach an ``infinitely'' long period for some energies. Moreover, if the energy is large enough this probability tends to unity \cite{hans2}.
Next, Takesue \cite{takesue} focused on the possible realization of statistical mechanics for reversible cellular automata. His studies concerned explicitly all class of rule in the one dimensional case, the Q2R being only a special case. However, the Q2R (90R in his terminology), is the analogue of an ideal gas of particles with speeds +1 or -1, a system that cannot reach equilibrium in practice. But, it is ergodic only in thermodynamical equilibrium. 
More recently, in Ref. \cite{goles}, one of us (SR) has studied numerically the irreversible behavior and the existence of a spontaneous transition from a non-coherent state to a coherent state.

 The present article is organized as follows, the Q2R model, as well as its main features and findings are presented in section \ref{model}. This section is sub-divided as follows: we briefly report the numerical studies of Ref. \cite{goles} in sub-section \ref{longtime}; the phase space  properties, in particular some results on the distribution of periods of the dynamics is reported in subsection \ref{PhaseSpace}; 
 and, finally, the scope of the paper is presented in sub section \ref{scope}.  
 Section \ref{masterEqn}, introduces the notion of a master equation for the statistical description of the dynamics. Finally, in Section \ref{Results} we provide some precise examples, where a coarse graining is realized, in order to get an adequate and tractable master equation. We provide an exhaustive validation of the technique and we discuss different coarse-graining over the phase space. Finally, Sec. \ref{Conclusions}  shows our conclusions.

 \section{The Q2R model.}\label{model}
For simplicity we shall consider a regular two dimensional lattice with $N=L^2  $ nodes, in which each node is only seen by its four closest neighbors (the von Neuman neighborhood), finally we use periodic boundary conditions.
Each node $k$ possesses a discrete value  $x_k$ that may take a value +1 or -1. The Q2R model, introduced by Vichniac \cite{vichniac}, is based upon the following two step rule:

$$
 x^{t+1}_k = x^{t-1}_k \, \phi\left( \sum_{i\in V_k} x^{t}_{i} \right),
$$
where the function $\phi$ is such that $\phi(s=0) = -1$ and $\phi(s) = +1$ if $s\neq 0$. In the sum $V_k$ corresponds to the von Neuman neighbor in the site $k$. The reversibility follows directly from the inverse relation $
 x^{t-1}_k = x^{t+1}_k \, \phi\left( \sum_{i\in V_k} x^{t}_{i} \right),$ which is the backward rule (notice that $  \phi\left( \sum_{i\in V_k} x^{t}_{i} \right)^2 =1 $ in all cases).

This two step rule may be naturally re-written as an one step rule by introducing a second dynamical variable \cite{pomeau84}:
\begin{eqnarray}
y^{t+1}_k & = & x^{t}_k \nonumber \\
 x^{t+1}_k &= & y^{t}_k\,  \phi\left( \sum_{i\in V_k} x^{t}_{i} \right).
 \label{q2r}
\end{eqnarray}
The rule (\ref{q2r}) is complemented with the initial condition $x_k^{t=0} $ and $ y_k^{t=0} $.

As shown by Pomeau \cite{pomeau84}, the energy
\begin{eqnarray}
E[\left\{x^t,y^t\right\}]  = -\frac{1}{2} \sum_{\left< i,k\right>}  x^{t}_k y^{t}_i,
\label{energy}
\end{eqnarray}
is preserved, $E[\left\{x^t,y^t\right\}] = E[\left\{x^{t=0},y^{t=0}\right\}] $ under the dynamics defined by the Q2R rule (\ref{q2r}). Moreover, the energy is bounded by $ -2N \leq E\leq 2 N$.

Despite the existence of an energy-like quantity, it is not possible to speak about a Hamiltonian discrete dynamic because, the variables $x^t$ and $y^t$ and the energy (\ref{energy}) are discrete quantities \cite{pomeau84}. Moreover, supported by the existence of a large number of periodic orbits (see next), it is believed that Q2R possesses a large number of other invariants, however, up to date, other conserved quantities are not known yet.

 \subsection{``Long-time'' dynamics of the Q2R cellular automata.}\label{longtime}

Numerical simulations of the Q2R model in two space dimensions  for large system sizes, {\it eg.} $N=256\times 256$, and random initial conditions, shows that the dynamics displays a fluctuating spatio-temporal pattern showing regions with states +1 and sectors with states $-1$, as well as, zones with chessboard-like pattern  \cite{goles}. The full patterns will be characterized by the global magnetization~:
\begin{eqnarray}
M(t)= M[\left\{x^t\right\}]  =  \sum_{k}  x^{t}_k ,
\label{magnetization}
\end{eqnarray}
Naturally, the function $M$ is restricted to the interval $-N\leq M\leq N$, and the available values of $M$ are ranged in uniform steps of $\Delta M =2$.

A detailed characterization of the evolution, as well as, of the fluctuations of the magnetization has been treated in detail in Ref. \cite{goles}. Briefly,  after a transient the average magnetization  depends mainly on the initial energy. If the energy is low, one sees that the average magnetization evolves slowly in time to an ``equilibrium'' state with an almost constant value plus weak fluctuations. For larger energies, the fluctuations enter to play an important role. One may observe that the system is in an almost stable state, but then suddenly jumps into a metastable state with zero average magnetization, and then jumps into an opposite magnetization state \cite{goles}. 

\begin{figure}[h!]
\begin{center}
  \includegraphics[width=7cm]{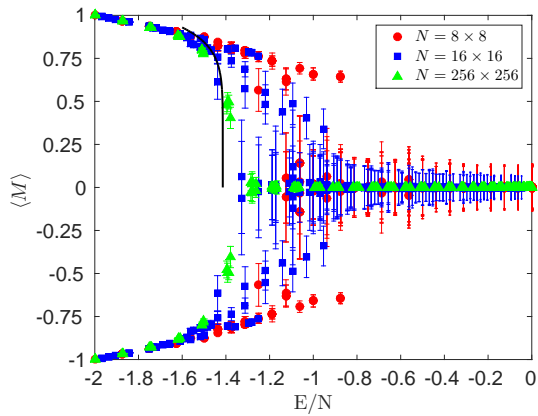}
\caption{ \label{MvsE}  Magnetization curves as a function of initial energy for three different system sizes: $N=8\times 8$, $N=16\times 16$, and $N=256\times 256$, as indicated in the inset. Each point corresponds to a different initial condition, in this case we sample different energies. As it can be noticed, there is a finite system dependence on the critical behaviour of the system, indeed the critical behavior disappear for small system sizes $N=8\times 8$, $N=16\times 16$, whereas for large system the magnetization curve reach a critical behavior. The continuous line represents the well known statistical mechanics calculation for the Ising model $M/N \approx 2^{5/16} (\sqrt{2} + E/N)^{1/8}$.}
\end{center}
\end{figure}

The plot of the temporal average of the global magnetization versus the energy is done in Fig. \ref{MvsE}. One sees that the magnetization spontaneously increases below a critical energy per site around  $E_c/N = - 1.4$, close to the critical energy of the Ising model $E_c/N = - \sqrt{2}$. \cite{onsager,yang}. Moreover, in Refs. \cite{hans,goles} it is compared the magnetization as a function of the internal energy of the system showing a close agreement with the numerical values.

\subsection{The phase space.}\label{PhaseSpace}

  The configurational space of all states is defined through all possible values of the state $\{x,y\}$. The resulting space is composed by the $2^{2N}$ vertices of a $2N$-dimensional hypercube. The smallest possible system corresponds to a $N=2\times 2$ lattice. In this case there are $2^{2\times 4}=2^8=256$ states and the phase space is a hypercube in dimension 8, however the dynamics is too simple, it does contain cycles of period 4 at most.
The phase space of a $4\times 4$ system is the largest possible one that can be studied exactly case by case.  In this case the system possesses $2^{2\times 16}=2^{32}$ states and it contains a rich variety of cycles (see Table \ref{Example4x4}). This case will be studied deeply as a good benchmark for conjectures in larger dimensional systems. 


As an example, from this case, it is observed that the total number of cycles, $n(T,E)$, of period $T$ and energy $E$, would be bounded by \cite{cycles}  
$$ n(T,E) < \frac{1}{T} 2^{2N} e^{-\alpha |E|} \sim e^{2 N \log 2 - \alpha |E|}.$$
From the data one has that for $3\times 3$ and $4\times 4$, $\alpha \approx 0.6$, but this value varies as the lattice size increases. Here one notices a dramatic difference among the cases $|E|$ greater or smaller than $E_c=\frac{2}{\alpha} \log 2$, if $|E|$ is greater than $E_c$ the probability to see a long period is exponentially small, but on the contrary, for $|E|<E_c$ this probability reaches the unity. Higher lattice sizes confirm this scenario but modify slightly the value of $\alpha$. This behavior is  consistent with the numerical simulations of Ref. \cite{hans2}.

\subsection{Scope of the paper}\label{scope}

Though the Q2R model is quite simple its dynamics is usually very rich as it has been documented extensively in literature. Moreover, this conservative and reversible system appears to behave as a typical macroscopic system, as the number of degrees of freedom increases, showing, in particular, a typical irreversible behavior, sensibility of the initial conditions, a kind of mixing, etc. It is believed that this Q2R is a good representation of an Ising model in thermodynamical equilibrium.

The phase space of the Q2R system of $N$ sites possesses $2^{2 N} $ states, which is partitioned in different sub-spaces of constant energies, which itself are partitioned in a large amount of smaller subspaces of periodic orbits or fixed points. Notice that, because the system is conservative, there are neither attractive nor repulsive attractors, all attractors are fixed points or cycles. 

\begin{figure}[h!]
\begin{center}
\centerline{ (a)  \includegraphics[width=7cm]{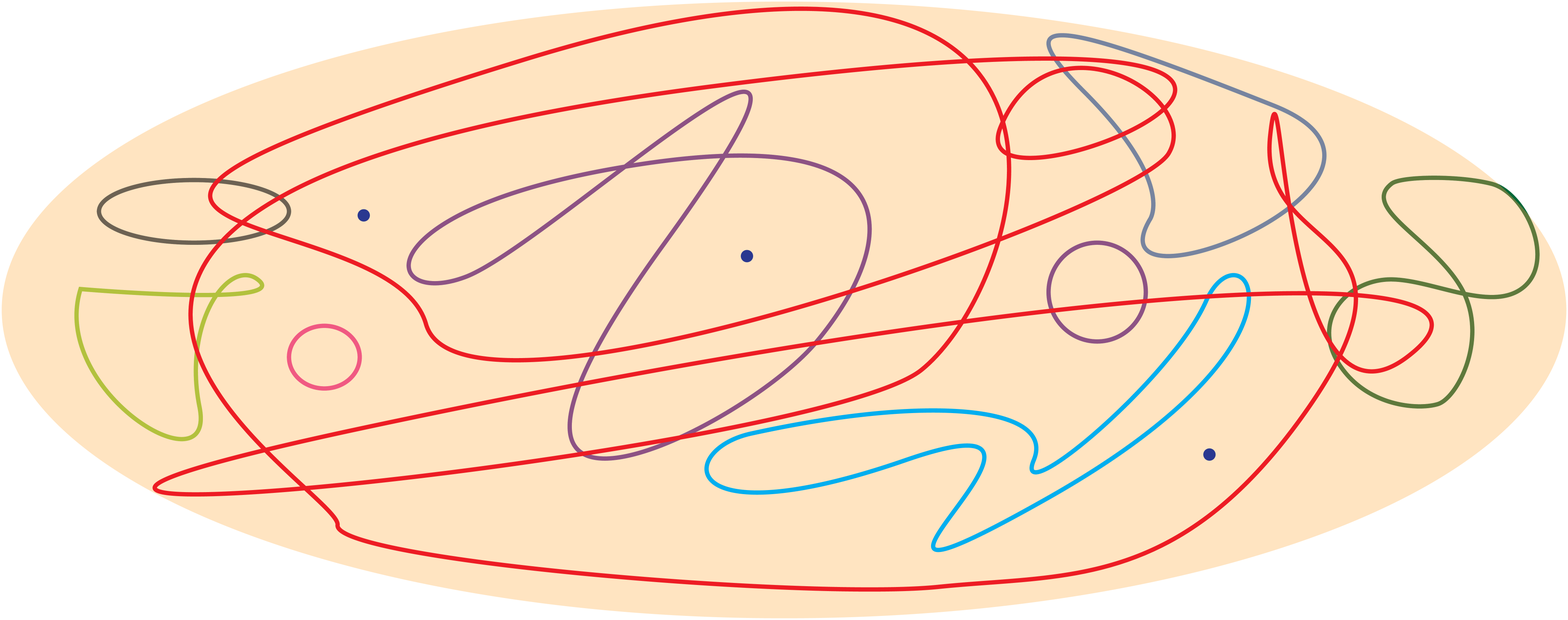}} \centerline{ (b)   \includegraphics[width=7cm]{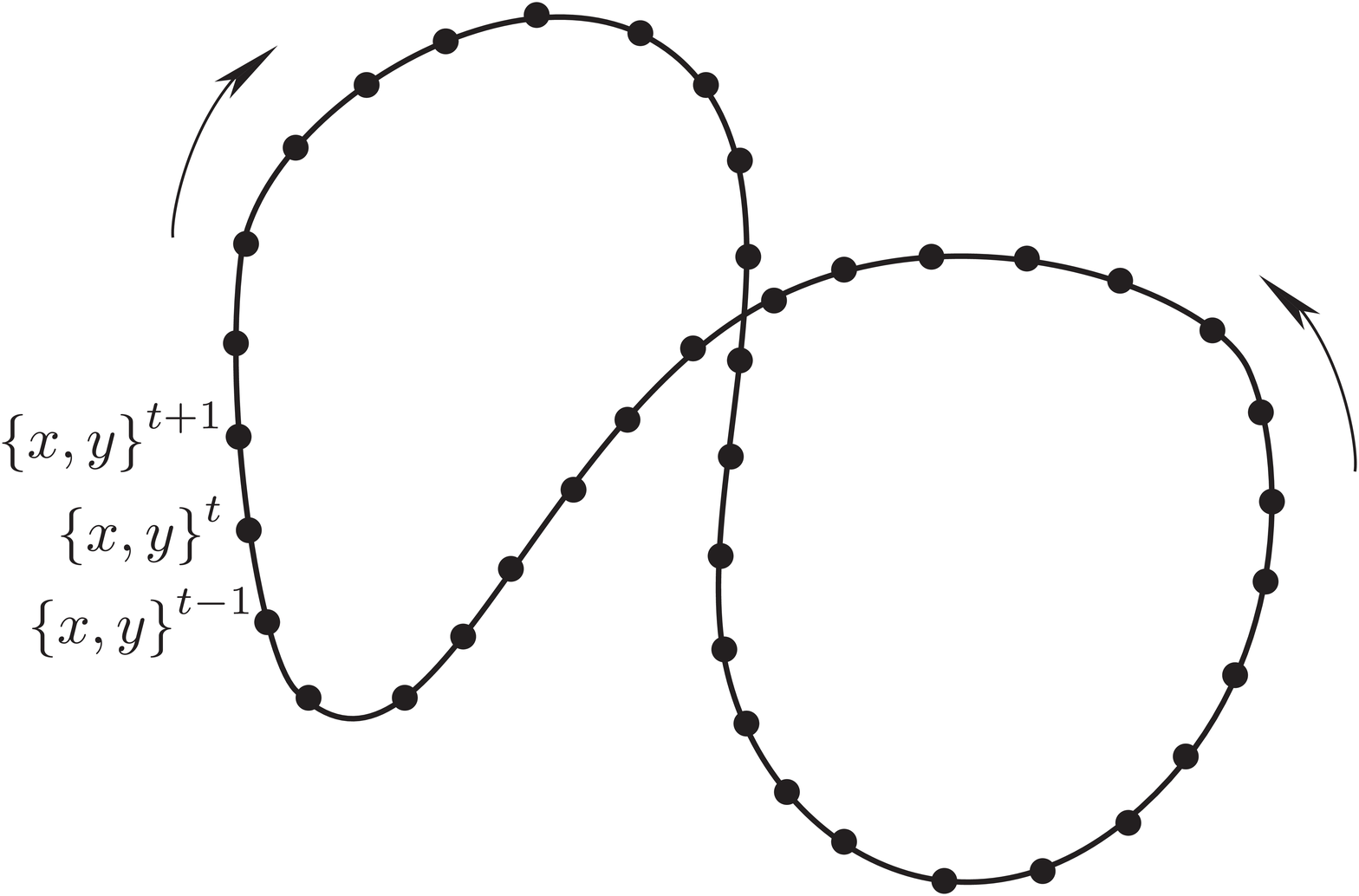} }
\caption{ \label{CycleDraft} (a) Scheme of a sub-space of constant energy composed by a number of cycles and fixed points. (b) Cartoon of a cycle of period $T$, for which the cycle is composed of $T$ states. }
\end{center}
\end{figure}

This feature of the phase space is schematized in Fig. \ref{CycleDraft} (a), where the constant energy subspace shares in principle many cycles and fixed points. An arbitrary initial condition of energy $E$ falls into one of these cycles, and it runs until it comes back to the initial configuration after a time $T$, which could be exponentially long and it displays a complex behavior (not chaotic stricto-sensu, see for instance \cite{grassberger}). More important the probability that an initial condition possesses such a complex behavior is finite \cite{hans2}. 
Moreover, Q2R  manifests ``sensitivity to initial conditions'', that is, if one starts with two distinct, but close, initial conditions, then, they evolve into very different cycles as time evolves \cite{goles}.  In some sense, an initial state explores  vastly the phase space justifying the grounds of statistical physics. 

In conclusion, the overall picture is~: although for a finite size system the deterministic automata Q2R possesses a periodic dynamics so it is not ergodic, there is a huge number of initial conditions that explore vastly the configurational space (this is particularly remarkably for initial conditions of random structure). Therefore, one expects that a master equation approach my be successful.

\section{Master equation}\label{masterEqn}


Given a set of initial conditions with a fixed energy $E$, the probability distribution  $\varrho^E_{t}(\{ x ,y\})$ evolves following a Perron-Frobenius like-equation
$$\varrho^{E}_{t+1} = {\mathcal L}^{E} \varrho^{E}_{t}$$
which, in principle, can be computed after the microscopic rule of evolution (\ref{q2r}).  Indeed ${\mathcal L}^{E}$ is easy to build: if the state $\{x,y\}_i$ at time $t$ goes to $\{x,y\}_k$, at a time $t+1$, then one sets the $(i,k)$ components to 1, that is ${\mathcal L}^{E}_{ik}=1$. Checking all available elements, $\Omega(E) $, for a given energy we built the huge,  $\Omega(E) \times \Omega (E) $, linear operator, ${\mathcal L}^{E}$. This matrix possesses a large number of blocks and zeroes revealing the existence of a large number of cycles in the Q2R model (In some sense, ${\mathcal L}^{E}$ is a kind of adjacency matrix of a graph, the graph being the total number of cycles shown for a given energy).

However, this description is impractical because of the typical magnitude of $\Omega(E)$. Therefore, the full phase space is reduced to a description using gross or macroscopic variables, namely the total magnetization of the system (\ref{magnetization}), instead of microscopic variables. Let us denote by $ \rho_t(M)$, the reduced probability distribution in terms of the variable $M$. The final  master equation will be of the form $\rho_{t+1}(M) = {\mathcal W}  \rho_{t}(M).  $
The linear operator, $ {\mathcal W}$, acts only in the subspace of constant $E$, but is spanned over arbitrary values of the magnetization.

As the original Perron-Frobenius  equation, $ {\mathcal W}$ depends explicitly of the Q2R rule and  it may be computed in principle. However, in practice, it is not possible because of the complex structure of the dynamics, in particular because of the existence of a myriad of different periods for a given $E$. Empirically, it seems that the probability to reach a long period is finite, therefore it is probable that any random initial condition would reach a extremely long cycle and the sequence of magnetizations would represent a  Markov chain $\{\cdots,M_{t-1},M_t,M_{t+1},\cdots\}$. 
Moreover, it is possible to reduce again the information via a coarse-graining partition of the possible values of the Markov chain. This partition is defined through a finite number of  sets of  no overlapping intervals:
$I_1 =[ -N , M_1 ), \,  I_2 =[ M_1,M_2 ),\,  \dots I_{K-1} =[ M_{K-2} ,\,  M_{K-1} ), \, I_K =[ M_{K-1},N ]$. In this context, we denote the original distribution function $\rho_{t}(M)$ by a discrete vector of dimension $K$, that is : $ \rho_{t}(M)\to  {\bm f}_{t} = (f_1,f_2,\dots f_K)$. 

Finally, we obtain a coarse-grained like master equation \cite{nicolis,et}
\begin{eqnarray}
{\bm f}_{t+1} = \hat W^\dagger {\bm f}_{t},
\label{MasterEqn}
\end{eqnarray}
where $\hat W$ is the probability transition matrix defined via the following conditional probability:$$w_{ik} = P(M_{t+1} \in I_i | M_t \in I_k)=  \frac{P(M_{t+1} \in I_i \cap M_t \in I_k)}{P(M_t \in I_k)}.$$
Here $M_{t}$ is at interval $I_k$ at time $t$, and  $M_{t+1}$ will be at interval $I_i$ at $t+1$. Finally, $\hat W^\dagger$ denotes the transpose of $\hat W$, and, the matrix $\hat W$  does not depend on time, which is a feature of a Markov process.

The coarse-graining method is schematized in the following Fig. \ref{Scheme}. 

\begin{figure}[h!]
\begin{center}
\centerline{\includegraphics[width=8cm]{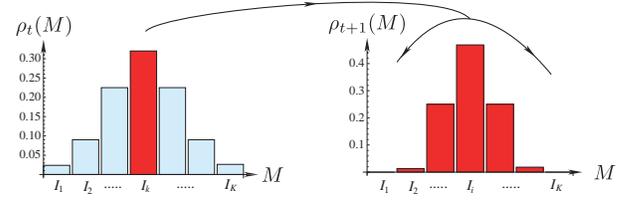} }
\caption{ \label{Scheme} The distribution $\rho_t(M) $ at a time $t$ is schematized in the left distribution. The fraction inside the interval $I_k$, is distributed, after the evolution into a new distribution $\rho_{t+1}(M)$ schematized in the left diagram. The normalized distribution provides the $i$-th element of the $i$-th column: $w_{ik}$.
}
\end{center}
\end{figure}
Important features of the master equation (\ref{MasterEqn}) are:

1) The probability vector ${\bm f}_{t}$ should be  positive and normalizable.  Let ${\bm 1} = (1,1,\dots 1)$ be a $K$-dimensional vector, then we set $ {\bm 1}\cdot {\bm f}_{t} =1.$ More important, because of  normalization,  $\sum_{k=1}^Kw_{ik} =1$,  one has $ \hat W \cdot {\bm 1}= {\bm 1}$. This implies that the probability is conserved under the evolution ${\bm 1}\cdot {\bm f}_{t+1} =  {\bm 1}\cdot \hat W^\dagger {\bm f}_{t}= {\bm 1}\cdot {\bm f}_{t} = 1.$

2) The Perron-Frobenius equation maybe solved exactly, provided an initial given distribution ${\bm f}_0$: ${\bm f}_{t} = (\hat W^\dagger )^t {\bm f}_{0}$.

3) Because of the Frobenius theorem,  it exists an eigenvalue which is one, $\lambda_1=1$, while others eigenvalues are inside the unitary circle $|\lambda_i|<1$ for $i>1$. 
Let ${\bm f}_{eq} $ be the Eigenvector associated with the Eigenvalue $\lambda_1=1$; this is an invariant vector ${\bm f}_{eq} = \hat W^\dagger  {\bm f}_{eq} $. 

4) In the following, we denote by ${\bm \chi}^{(i)}$ the eigenvectors of $\hat W^\dagger$ corresponding to $\lambda_i$. Naturally one has ${\bm \chi}^{(1) }\equiv  {\bm f}_{eq}.$ 

5) The existence of an equilibrium state: $\lim_{t\to\infty} {\bm f}_{t} = {\bm f}_{eq}$.   

6) Because all elements in the $W$-matrix are positive, any non negative initial distribution remains non negative.

\subsection{The Chapman-Kolmogorov condition and time reversal symmetry.}\label{ChapmanKRelations}

One may wonder if one realizes the same process but instead to look at the system at times $t$ and $t+1$, one looks at time $t$ and $t+2$ or more generally at $t$ and $t+\tau$. Let us call $\hat W^{(\tau)}$ the resulting probability transfer matrix after $\tau=\tau_1 + \tau_2$ iterations, therefore it is easy to show that this matrix should verify the Chapman-Kolmogorov or compatibility condition \cite{nicolis}, 
\begin{eqnarray} \hat W^{(\tau)}=  \hat W^{(\tau_1)} \cdot  \hat W^{(\tau_2)} . \label{ChapmanK} \end{eqnarray}

In particular, for $\tau_1=\tau_2=1$ one should satisfy
$$ \hat W^{(2)}=  \hat W\cdot  \hat W=  \hat W ^2 ,$$
which is true for conditional probabilities, because of the relation 
$P(M_{t+2} \in I_k | M_t \in I_i)= \sum_j P(M_{t+2} \in I_k | M_{t +1}\in I_j) P(M_{t+1} \in I_j| M_t \in I_i),$ 
which is equivalent to the right hand side. 

Other compatibility conditions are 
\begin{eqnarray}
 \hat W^{(3)}& = &  \hat W^{(2)}\cdot  \hat W, \nonumber \\
\hat W^{(3)} & = & \hat W \cdot  \hat W^{(2)}, \nonumber \\
\hat W^{(4)} & = &\hat W^{(2)} \cdot  \hat W^{(2)}, \nonumber \\
\hat W^{(4)} & = & \hat W \cdot \hat W^{(2)} \cdot  \hat W, \quad {\rm etc.}\nonumber
 \end{eqnarray}


 

Finally, le us state an important result due to Pomeau \cite{pomeau82}. The $K$-time correlation functions imposes some restrictions on the $W$-matrix. 

Because of the time reversal symmetry, for all indices $i_1,i_2,\cdots i_K = \{ 1,2,\cdots K\}$, the symmetry relation 
\begin{equation}
w_{i_1i_2} w_{i_2i_3} \cdots w_{i_{K-1}i_K} w_{i_Ki_1} = w_{i_1i_K} w_{i_K i_{K-1}}   \cdots w_{i_3i_2}  w_{i_2i_1}\label{pomocondition}\end{equation}
must be satisfied.

In the following we apply this coarse graining approach to compute the probability transfer matrix for some particular cases.

\section{Specific computation of the probability transition matrix in various situations.}
\label{Results}
 In this section we shall apply the coarse graining approach to the Q2R dynamics in the case of small lattice size. In Ref. \cite{chine} we have fairly explored the computation of the probability transition matrix, in particular, in the case of extended systems ($N=256\times 256$), however in this case the cycles are usually huge, therefore this general approach is not really complete. In this sense, we focus our effort in treating moderate system sizes, namely $N=4\times 4$, $N=8\times 8$, $N=16\times 16$, having all of them tractable cycles. 

\subsection{The robustness of the methodology.}

In general for a system of small size, one is able to find some cycles for a given energy.  Building a time series for the magnetization $\{ M(t)\}= \{ M_1,M_2,\cdots M_T\}$, then one defines a partition on the possible values of the magnetization, as explained in a previous Sec. \ref{masterEqn}. In the cases considered here, it is always possible to use the finest possible partition, that is, for the exact available values of the magnetization (something impractical in large systems). In this case the partitions are composed by a set of $N+1$ ($N$ is assumed to be even) well defined values $ M=\{-N, -N+2,-N+4,\cdots , N-2,N\}$. That is for $4\times 4$ the partition has a maximum of 17 elements, for $N=8\times 8$ there are 65 elements, and for $N=16\times 16$ the partition possesses a maximum of 257 elements.


A first result concerns the equivalence of the probability density function of magnetization obtained via the time series of the magnetization and the equilibrium distribution resulting from the eigenvectors of the probability transition matrix $\hat W$.  Hence, the results arising from temporal averages and transition probability matrix in the configurational space are consistent themselves. This fact ensures a first validation of the method. However, the  probability transition matrix provides extra information of a system, among them, the non-equilibrium properties, given by the spectrum of $\hat W$.

Next, we shall describe the methodology for the case of a $16x16\times 16$ lattice size for an orbit of $E=-292$ and a period $T= 43115258$.
The probability transition matrix $\hat W$  is constructed following the steps of the previous section \ref{masterEqn}. But first, we shall verify that the master equation does not strongly depend on the length of the time series for the magnetization. It is important to underline that we think that this is a crucial step, because it allows us to compare explicitly the dependence of the results on the partial length of the cycles. Something not possible in larger systems, because in these cases we shall never be able to build the complete period for the time series.

\begin{figure}[h!]
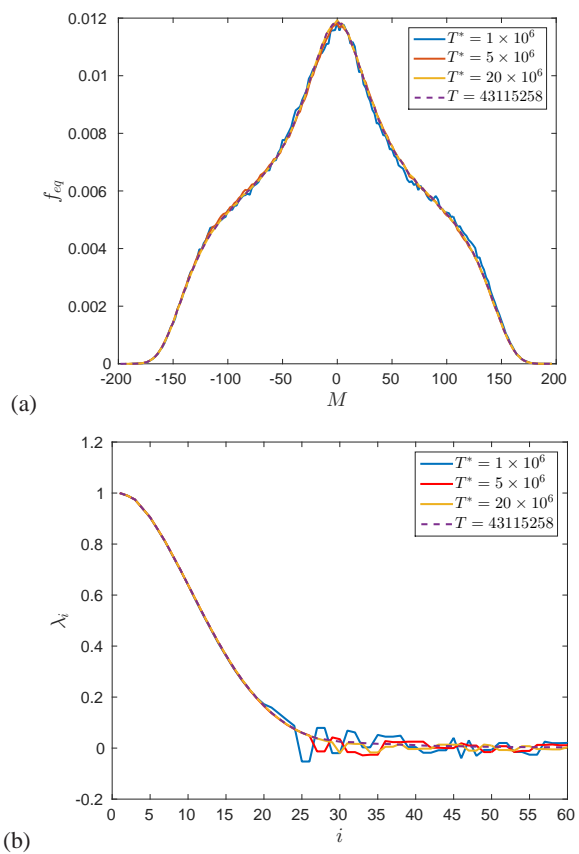

\begin{center}
 \centerline{   (a)  \includegraphics[width=7cm]{feq_N_16x16_E_292_T.eps}  } \medskip \centerline{ (b) \,\, \includegraphics[width=7cm]{Lamb_N_16x16_E_292_T.eps}  }
\caption{ \label{Method} (a)  Plot of the equilibrium distribution ${\bm f}_{eq} $ for the case of a $16\times 16$ system with $E=-292$ ($E/N \approx 1.14$) and a cycle of a period $T=43115258$. The computation of ${\bm f}_{eq} $ is compared with shorter sequences of the same time series of length  $T^* = 10^6 ,\, 5\times 10^6,\, \& \,  20\times 10^6$. (b) The set of 257 eigenvalues of the $\hat W$-matrix for the same conditions of (b). Naturally,  $\lambda =1$ is present by the definition of the probability transfer matrix $\hat W$.
}
\end{center}
\end{figure}

To test that, we shall use again the finest partition. In this case the transition matrix has dimension $257 \times 257$ (so we shall not provide them explicitly) and we shall characterize it  by the equilibrium distribution, and the full set on eigenvalues of $\hat W$.  Fig. \ref{Method} (a) and (b) plot the equilibrium distribution ${\bm f}_{eq} $ and the set of 257 eigenvalues for the same sequence, $\{ M(t)\}$, but for four different lengths of the time series: $T= 43115258$ (the complete cycle), and $T^* = 10^6 ,\, 5\times 10^6,\, \& \,  20\times 10^6$ (partial sequences). 
 Visually one sees that there is not substantial difference among the different values of $T^*$. Moreover, the following Table \ref{TableTest} compares quantitatively the mean square difference measuring $Q_1 = || {\bm f}^{*} -{\bm f}_{eq} ||^2/K$ and $Q_2 =\sum_{i=1}^K|\lambda_i -\lambda_i^*|^2/K$. Here $K$ is the number of partitions.

\begin{table}[h!]
\centering
 \begin{tabular}{|c||c|c| }
  \hline  
   \hline
$T*$&$Q_1$& $Q_2$ \\ \hline 
$10^6$ & $3.95\times 10^{-5} $& $0.0038 $ \\ \hline
$5\times 10^6$ & $3.91\times 10^{-5}$&  0.0020   \\ \hline
$20\times 10^6$ & $3.84\times 10^{-5} $& $0.0002$ \\ \hline
 \end{tabular}
  \caption{\label{TableTest} Error estimation of the equilibrium distribution and the spectral decomposition of the $\hat W$ matrix for different lengths of the time series.   }
\end{table}

 We have noticed that the coarse-graining approach applied to the full cycle with period $T$ presents an important characteristic. Namely, the eigenvalues of the $\hat W$ matrix are real numbers. However, as we apply the same approach to a partial sequence of the same cycle of a length less than $T$, some eigenvalues become complex values (typically located near the origin in the complex plane).  An  important consequences of the present work is the following conjecture: For complete cycles the probability transfer matrix possesses real eigenvalues, but for incomplete sequences the matrix does not possess, in general, only real eigenvalues. This subject will be matter of a future research.

Finally, it is important to compare the results for different partition sizes. 
First, we compute the equilibrium distribution for three partitions sets. More precisely, for a $8\times 8$ system evolving by Q2R at $E=0$ in a periodic orbit of  $T=672018$.




 
\begin{figure}[h!]
\begin{center}
 \centerline{(a)\,\, \includegraphics[width=7cm]{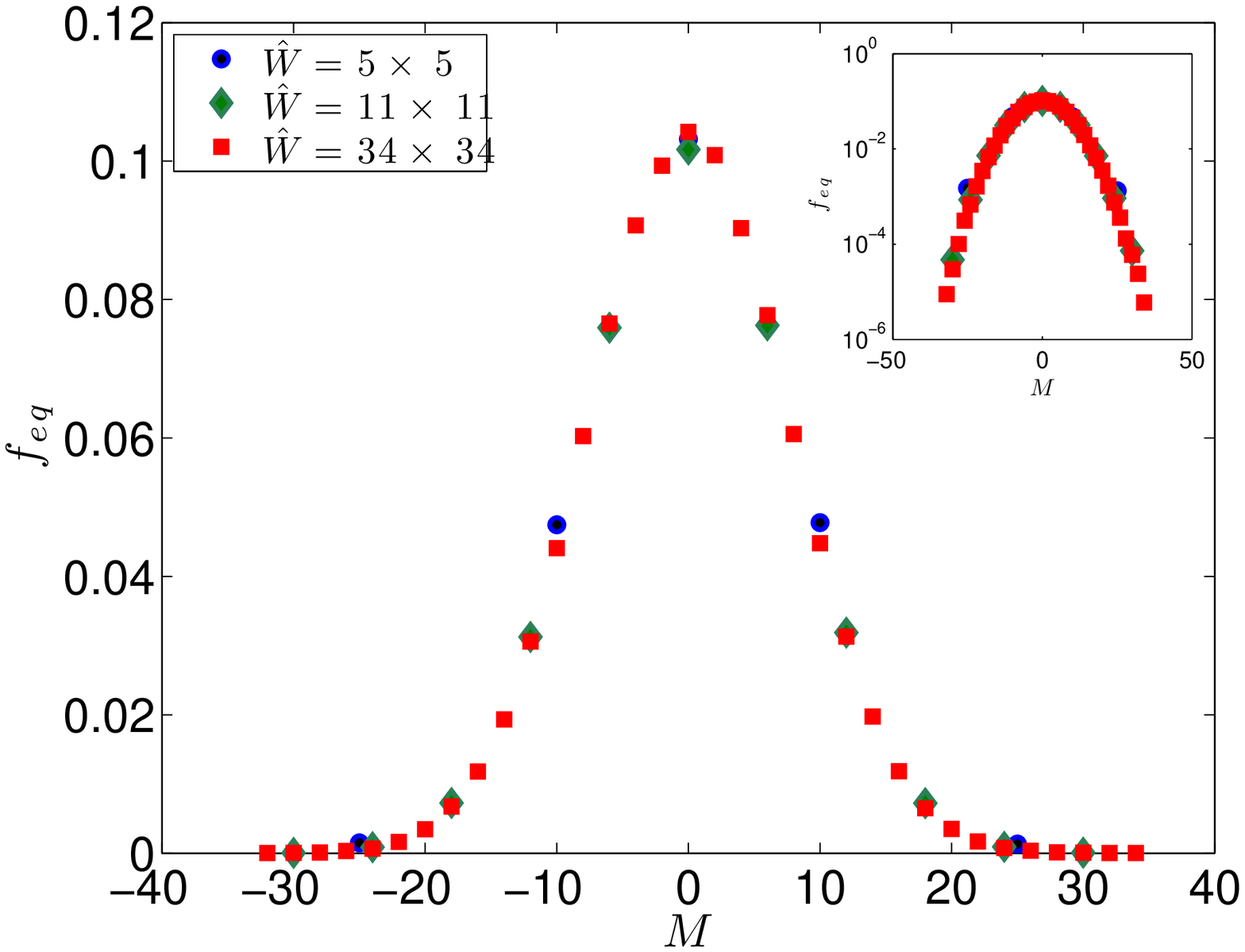} } \medskip \centerline{   (b) \,\,    \includegraphics[width=7cm]{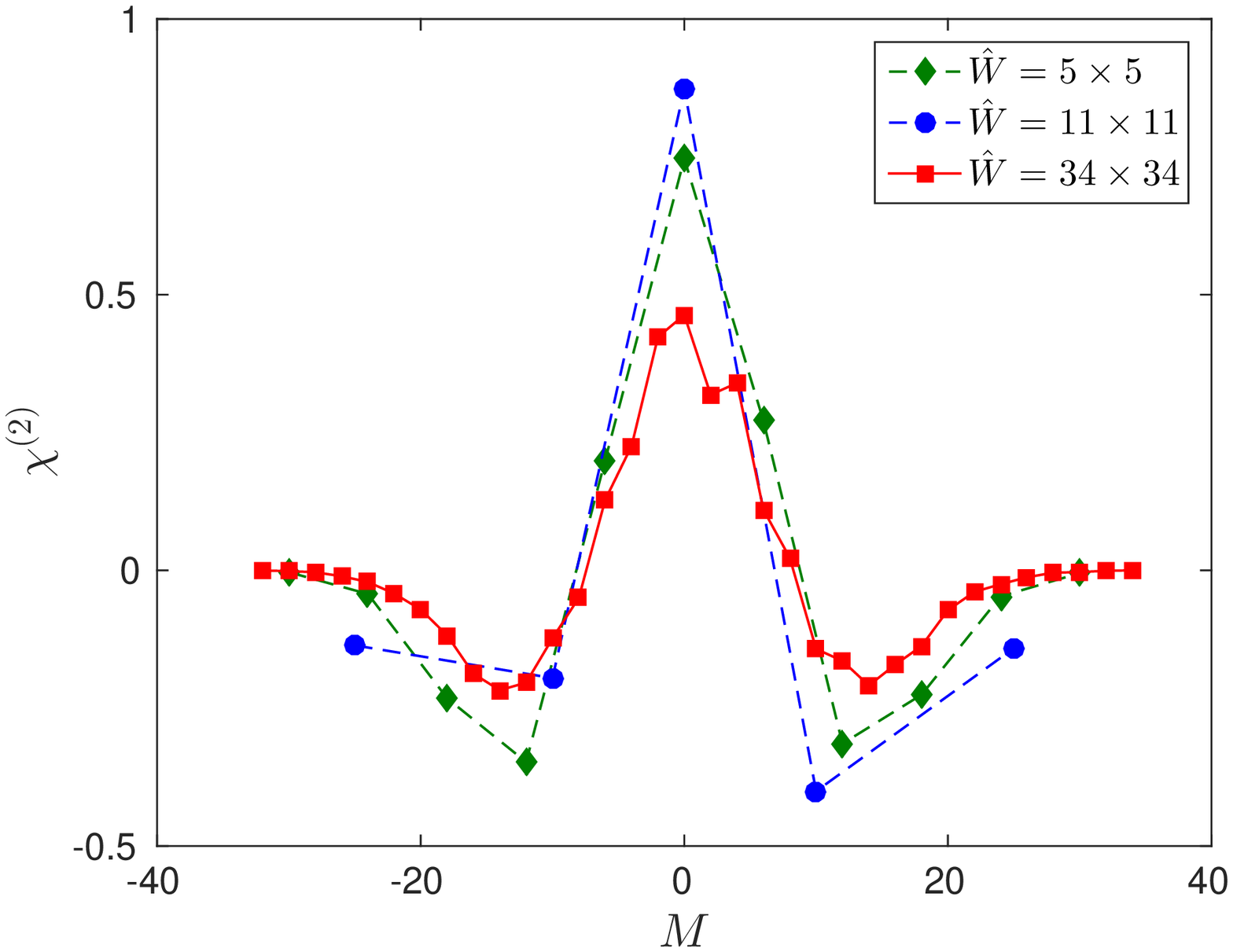}    }
\caption{ \label{Method88} (a) Plot of the equilibrium distribution $f_{eq}$ vs $M$ for a $8\times 8$ system with $E=0$ and a cycle of a period $T=672018$ for three different partitions of the magnetization values. The plot shows that all distribution functions lies under the same curve.  The inset shows the parabolic behavior in magnetization which after a fit  reads $ \log f_{eq}= -M^2/ 116$. (b) Plot of the second eigenmode ${\bm \chi}^{(2)}$ corresponding to the eigenvalue closest to the unit circle. One notices that all partitions produce similar results.
}
\end{center}
\end{figure}

Figure \ref{Method88} (a) compares the three different coarse-graining partitions (5, 11 and 34 elements of the partition). Despite the evident differences among the coarse and the fine graining partitions, one notices that both partitions possess the same accurate behavior of the equilibrium distribution. Moreover, Figure \ref{Method88} (b) compares the second eigenmode ${\bm \chi}^{(2)}$ without any substantial difference among the partitions.

In the following we summarize the methodology for the cases of $4\times 4$, $8\times 8$, and $16\times 16$, in all cases the full cycles were considered, and we provide the finest partition. 


\subsection{Exact calculation for various lattices.}

We have studied in detail the case of a $4\times 4$ periodic lattice, because the phase space possesses $2^{32} \approx 4\times10^9$ distinct configurations  and the calculations are exactly realizable in a computer up to end, therefore it shows explicitly the method.  It is shown that the coarse graining approach is fully applicable in the $4\times 4$ lattice case. We used different partition getting a well defined probability transfer matrix $\hat W$. 
The Table \ref{Example4x4}, reproduced in Appendix \ref{App1}, summarizes the calculations and main characteristics for various energies. 

Next, we shall explore few cycles for larger systems ($8\times 8$ and $16\times 16$). The cycles in these cases may be as long as desired for any practical purpose, so that the equilibrium distribution is solved with enough precision.

In the case of $8\times 8$, for various energies and the finest coarse graining, as a sake of brevity, we omit explicitly the plots of  the first eigenvector, ${\bm f}_{\rm eq}$, as well as the eigenvalues, because they are similar to the one of the $16\times 16$ lattice case. 

\begin{figure}[h!]
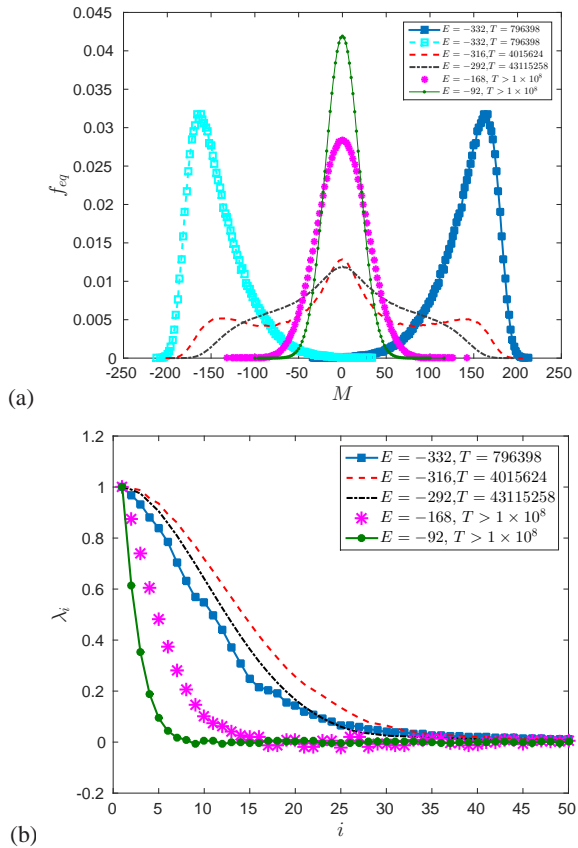

\begin{center}
\centerline{ (a)\,\, \includegraphics[width=7cm]{feq_N_16x16_all.eps}  }  \medskip  \centerline{ (b)\,\, \includegraphics[width=7cm]{valorP_N_16x16_all.eps}}
  \caption{ \label{Eigenvalues1616} (a) Equilibrium distributions, ${\bm f}_{\rm eq}$, for the case of a $16\times 16$ system, and for the energies and periods: $E= -332$ and  $T=796398$, $E= -316$ and  $T=4015624$,  $E= -292$ and  $T=43115258$. We also considers $E=-168$ and $E=-92$ with periods larger than $T>10^8$. (b) Plot in semi-log scale, $\log \, {\bm f}_{\rm eq} $ vs. $M$ confirming a typical exponential behavior of the inset of Fig. \ref{Method88}-a. (c) Eigenvalues of the $W$-matrix showing the existence of long-wave relaxation properties. }
\end{center}
\end{figure}

The case of a $16\times 16$ system presents the most accurate equilibrium distributions found in the current research. The fluctuations around the distribution are small, and the eigenvalues seems to form a continuous spectrum (the difference among two consecutive eigenvalues is small). We have also explored  a wide range of energies. 
The rank of the matrices (that is the finest partition) are $K= 122$ for  $E= -332$; $K=205$ for $E= -316$; $K= 197$ for $E= -292$; $K=129$ for  $E=-168$; and $K= 101$ for $E=-92$.
The equilibrium distribution, as a function of the magnetization, is plotted in Fig. \ref{Eigenvalues1616}-a. Similarly the spectral decomposition is shown in  Fig. \ref{Eigenvalues1616}-b.

In Fig. \ref{Eigenvalues1616}-(a) one notices in the case of large energies, say $E=-92$ and $E=-168$, the equilibrium distribution function is symmetric, under the change $M\to -M$, however as the energy decreases one sees that for the lowest energy, $E=-332$, it appears a spontaneous symmetry breaking, so that the equilibrium distribution is not anymore an even function.  The equilibrium probability may manifest a positive or negative magnetization (switching from one case to the other by changing  the initial condition via the transformation $\{x,y\} ^{t=0}\to \{-x,-y\}^{t=0}$).
Moreover, the energy $E= -316$ case shows an equilibrium probability density function that manifests bi-stability. Indeed, these bi-modal distributions possess three peaks, one at $M=0$ and two other at $M=\pm M_0\neq 0$. Finally, the width of the probability density functions increases near the transition energy.

Fig. \ref{Eigenvalues1616}-(b) shows the spectral distribution of  the probability transfer matrix that defines the master equation. Already in a lattice size of $16\times 16$ one observes that the spectral distribution is almost continuous. One notices that the energies $E= -316$ and $E= -292$ possess the largest eigenvalues for a given index $i$. That means that probably the largest eigenvalues occurs near the critical energy.   

It is interesting to remark that the non-equilibrium is governed by the eigenvalues near the unity. The non-equilibrium features behave as slow modes. In the current case one has ${\bm f}_t =  \sum_{i=1}^K \alpha_i \lambda_i ^t {\bm \chi}^{(i)}.$ Defining $\sigma_i =- \log\lambda_i$, one obtains the usual slow mode relaxation.  
Moreover, the global behavior of the eigenvalues closest to the unity, represents the transport coefficients \cite{chine}.
 Fig. \ref{Eigenvalues1616}-(b) indicates that $\lambda_i \approx 1 - \gamma i$, something that suggests that the  non-equilibrium features are governed by a Fokker-Planck kind of equations. The behavior of the eigenvector agrees also qualitatively with this picture (see \cite{chine} for more details).

\subsection{The Chapman-Kolmogorov conditions.}
We have checked the Chapman-Kolmogorov relations for the case of Q2R in a $16\times 16$ lattice for the case of $E= -292$ and a periodic orbit of $T=43115258$. We have built five different probability transfer matrices $\hat W^{(\tau =1 )}, \cdots , \hat W^{(\tau =5 )}$(See Sec. \ref{ChapmanKRelations} for the definition of $\hat W^{(\tau )}$).

\begin{figure}[h!]
\begin{center}
\centerline{ (a) \, \includegraphics[width=7cm]{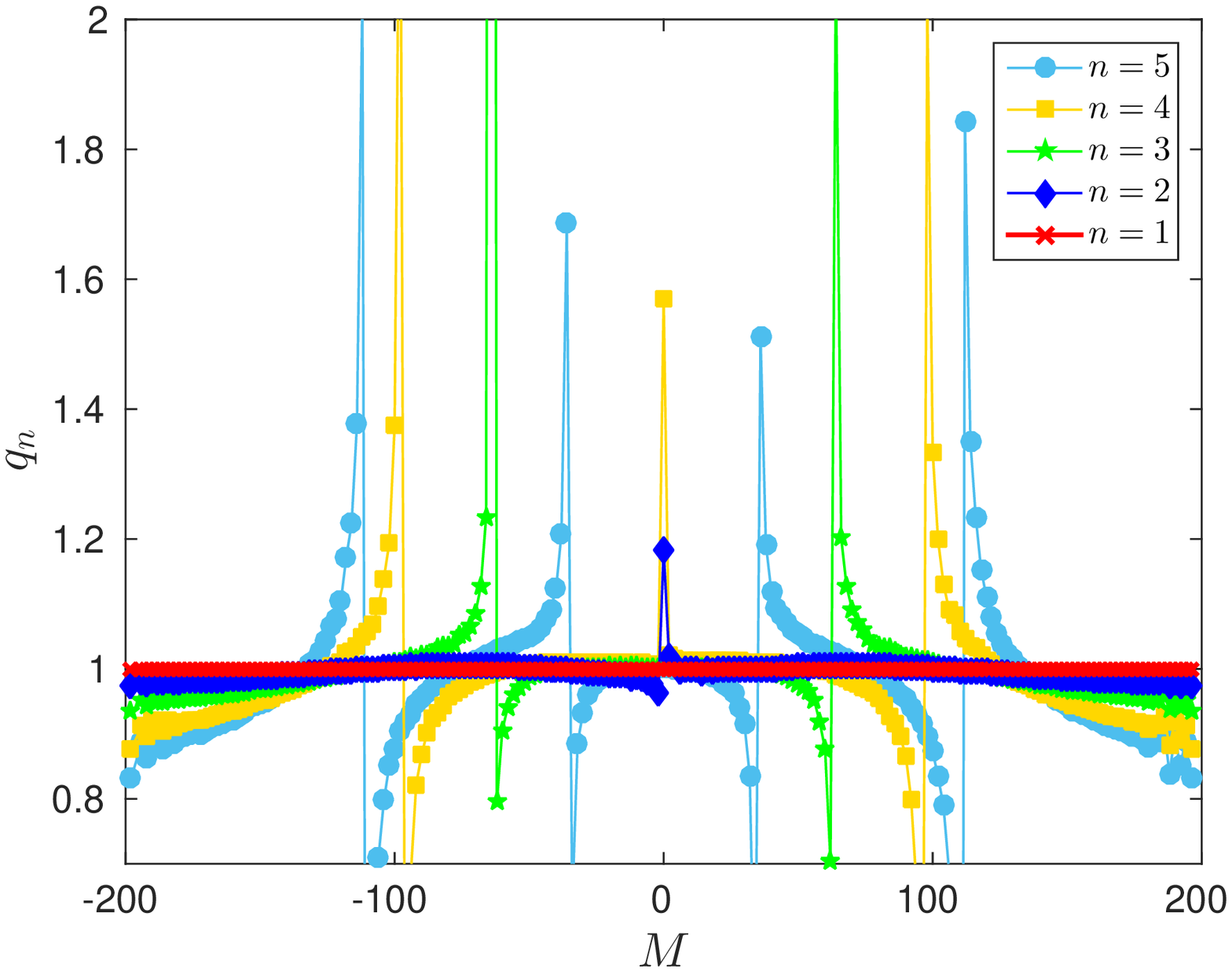} } \medskip
\centerline{(b)\,   \includegraphics[width=7cm]{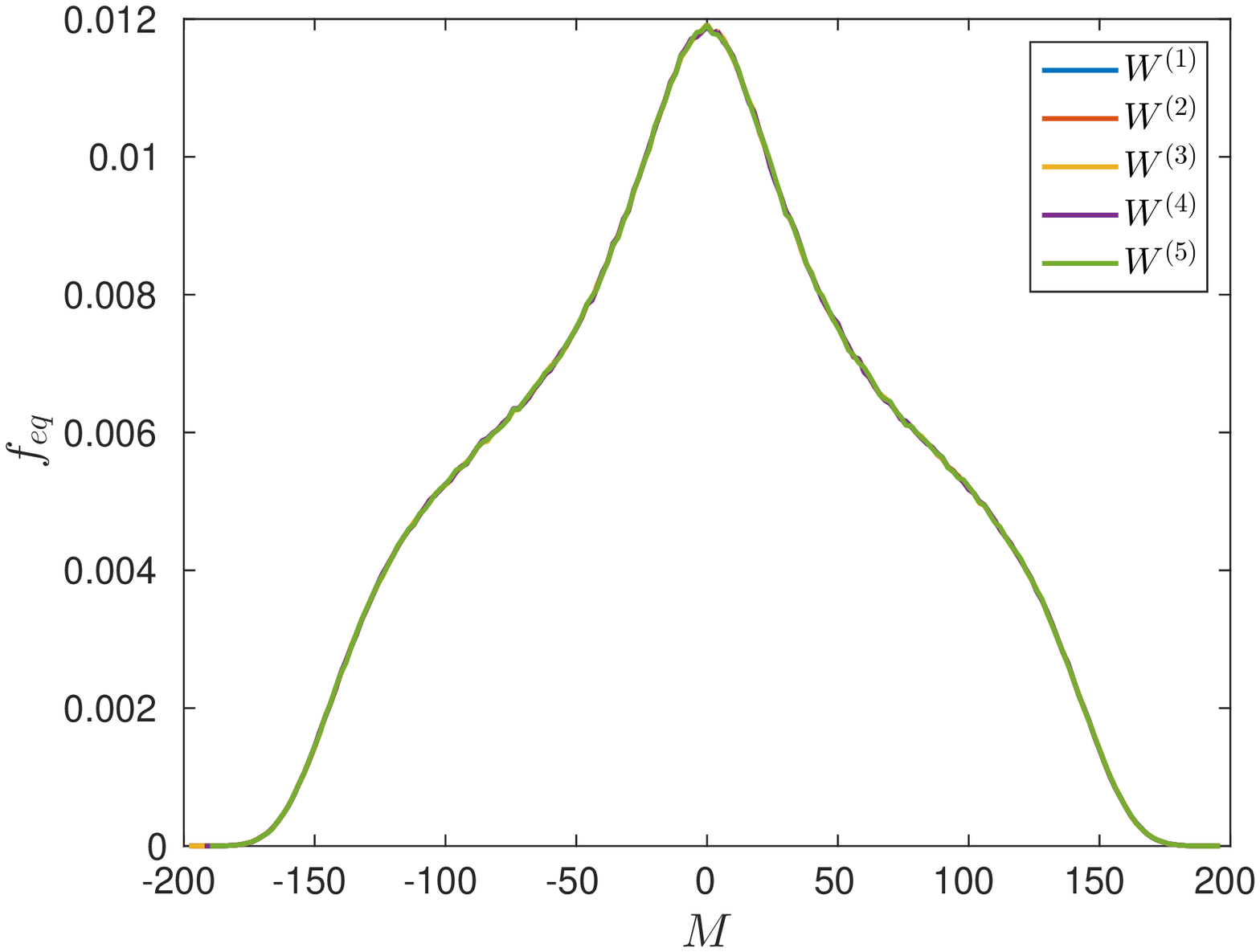} } \medskip
 \centerline{(c) \,  \includegraphics[width=7cm]{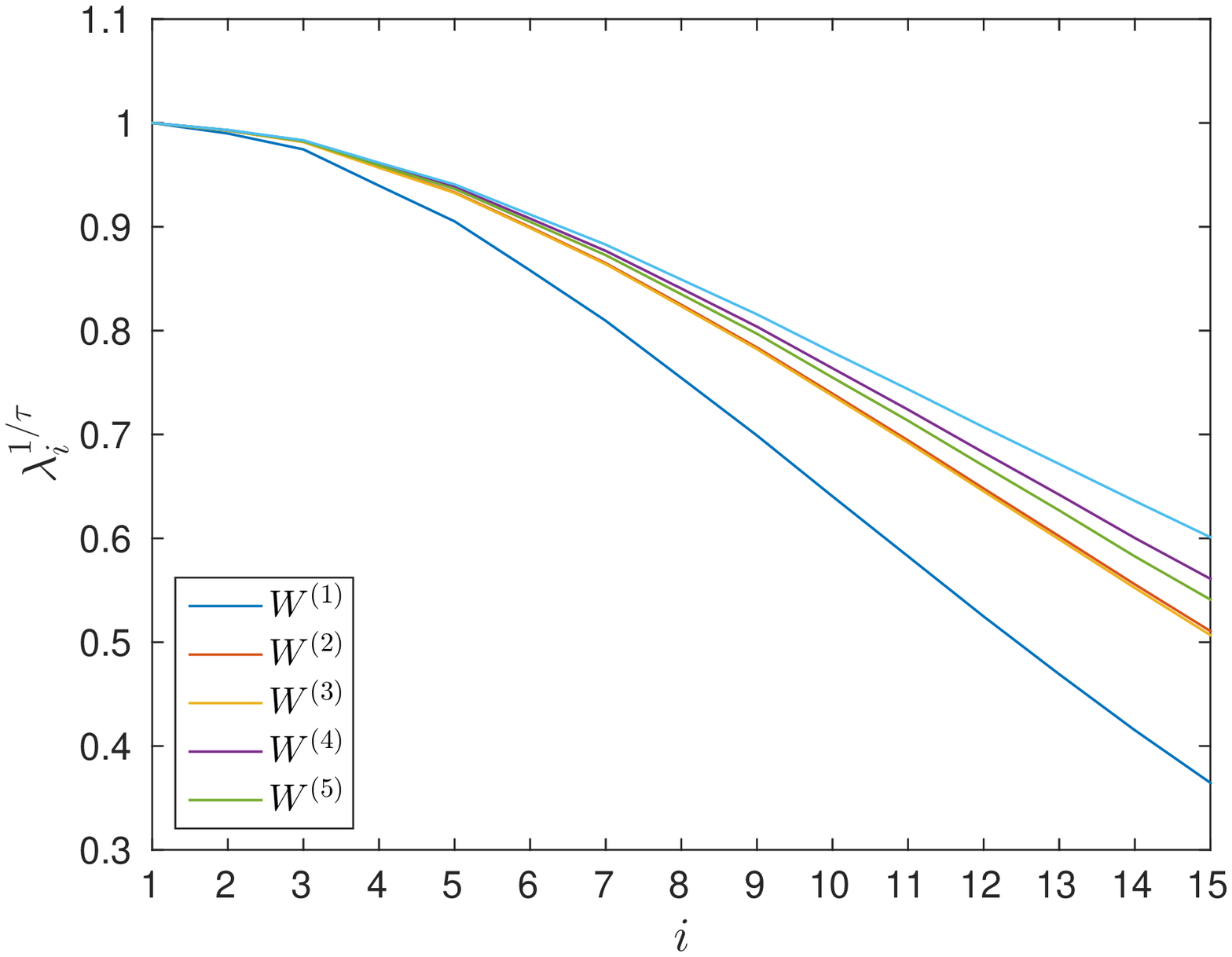}}
  \caption{ \label{CheckChapmanK} (a) Plot of the ratio $q_n$ for five eigenmodes. (b) Equilibrium distributions, ${\bm f}_{\rm eq}$, for the case of a $16\times 16$ system, and for the energy $E= -292$ and  $T=43115258$.  (c) Eigenvalues of the $W$-matrix showing the existence of long-wave relaxation properties. }
\end{center}
\end{figure}

 First, we compare the matrices
$\hat W^{(\tau =2 )}$ and $\hat W^{(\tau =1 )}\cdot \hat W^{(\tau =1 )}$, both of a rank $197\times 197$, computing the distance among the matrices  $\hat W^{(\tau =2 )}$ and $\hat W^{(\tau =1 )}\cdot \hat W^{(\tau =1 )}$, via the usual distance (the squared indicates the square of a matrix)  $$d=\frac{1}{K^2}{\rm Tr}[(\hat W^{(\tau =2 )}-\hat W^{(\tau =1 )}\cdot \hat W^{(\tau =1 )})^2].$$
In the current case, the matrices are similar up to $d= 5.81\times 10^{-6}$. 
More quantitatively, we look  how good are the eigenvectors of different matrices, namely $\hat W^{(\tau =2 )}$ and $\hat W^{(\tau =1 )}\cdot \hat W^{(\tau =1 )}$. To do that, we compute the ratio among the $n$-{\it th} eigenvectors of the afore mentioned matrices, that is $$q_n = \frac{{\chi^{(2)}}_n}{{\chi^{(1)}}_n },$$ where ${\chi^{(2)}}_n$ and ${ \chi^{(1)}}_n$ are the $n$-th eigenvector of the matrices $\hat W^{(\tau =2 )}$ and $\hat W^{(\tau =1 )}$. This quantity is plotted in Fig. \ref{CheckChapmanK}-(a). One notices that  $q_n\approx 1$ almost for all values of magnetization, but it has an anomalous behavior near the nodal points of the eigenvector ${\chi^{(1)}}_n $. In general the agreement of all this eigenvectors is satisfactorily.

Next we check, the Chapman-Kolmogorov relations written in Sec. \ref{ChapmanKRelations}, comparing the spectral properties of both matrices, namely the set of eigenvectors and its eigenvalues.  

As it can be seen in the Fig. \ref{CheckChapmanK}-(b) the equilibrium distribution ${\bm f}_{eq} $ matches perfectly for different values of $\tau =\{1,2,3,4,5\}$. This proves that the equilibrium configuration,  ${\bm f}_{eq} $, is well an invariant of the dynamical system. However, non-equilibrium properties do depend on the sampling time, $\tau$. Indeed, the eigenvalues corresponding to different probability transfer matrices do depend on the choice of the parameter $\tau$.  This is not a surprise, because it is expected that the eigenvalues, $\lambda^{(\tau)}_i$, of $\hat W^{(\tau  )}$ should scale as $\lambda^{(\tau)}_i=\lambda_i^\tau$, where $\lambda_i$ are the set of eigenvalues of  $\hat W^{(\tau =1 )}$. This scaling is shown in Fig. \ref{CheckChapmanK}-(c) indicating an anomaly because it does not work for the case $\tau=1$, but the scaling works well for higher $\tau$. This deserves more  careful study.

\subsection{Pomeau's reversal symmetry relation.}

 According to Pomeau  \cite{pomeau82}, the microscopic  time reversal symmetry imposes the symmetry relation (\ref{pomocondition}).
For a probability transition matrices of rank $K$, it is possible to verify that  there are  $K^K$ different required conditions (\ref{pomocondition}). 
Therefore, the verification of this condition is possible only for a moderate matrices rank, $K$. All probability transfer matrices written in Table \ref{Example4x4} of Appendix \ref{App1} satisfy the Pomeau's reversal symmetry relation (see the ``Reversible'' row in the table).

For larger $\hat W$ matrices, say $K>9$, we have not checked Pomeau's relation because it bemuse a cumbersome numerical calculation.

\section{Discussion}\label{Conclusions}
The basic properties of the Q2R cellular automata, namely, its formal reversibility and the existence of a conserved energy suggests that the Q2R could be a good benchmark to test ideas of statistical mechanics. Importantly, the reversibility is not conditioned by any approximative numerical algorithm. 
The Q2R model possesses a rich dynamics characterized by a huge number of invariants that partitioned the phase space in terms of the conserved energy and a huge number of periodic cycles. Although, in moderate system size the periods are huge numbers \cite{hans2}, in small lattices these cycles may be fully characterized.

We have introduced a coarse-graining approach, that allows us to write a coarse-grained master equation, that characterizes the equilibrium and non-equilibrium statistical properties of the system. We overview the methodology and test the consistency of the results in different lattice size. We can see that if the partitions are well done, this coarse graining technique is a powerful tool to reduce the information of whole system in a tractable probability transfer matrix which simplify the original master equation. A first central property of this matrix, is the existence of an invariant probability distribution which agrees for different coarse-graining procedures. Secondly, we compute the spectral decomposition of the  probability transfer matrix that characterizes the non-equilibrium properties of the system. We conjecture that the probability transfer matrix   possesses real eigenvalues, if it was built using the complete periodic sequence, but for incomplete sequences the matrix does not possess, in general, only real eigenvalues. Third, we check that validity of the Chapman-Kolmogorov relation, as well as, the time reversal symmetry conditions. In many situations the methodology satisfies the requirements and provides a complete statistical description of the system, but some discrepancies appears that deserves caution. 
This study provides a systematic approach to reduce the phase space to the pertinent macroscopical variables reducing the information up to a  handily master equation.

The authors acknowledge Professor Enrique Tirapegui to bring our attention to the coarse-graining methodology for reversible systems and for his participation in early stages of this work. The authors acknowledge F. Mora for valuable comments to this work. F.U. thanks CONICYT-Chile under the Doctoral scholarships N $21140319$, and S.R. thanks a CONICYT Basal-CMM grant. Also, we thank to Fondequip AIC-34.

\appendix

\section{EXAMPLES}
\label{App1}
\subsection{Exact calculation for the $4\times 4$ lattice.}\label{App:Case44}

Consider the case of a $4\times 4$ periodic lattice. Though the Q2R dynamics is extremely simple, the calculations are exactly realizable up to end and for all configurations, thus it may explicitly explain the method. The phase space possesses $2^{32}$ distinct configurations which may be computed directly. The energy takes possible values ranging from  $-32\leq E\leq 32$. We have characterized few special cases with energies and periods distributed uniformly over the all possible values: $(E,T) = (-24,6)\, (-22,10)\,(-18,54)\,(-8.270),\, (-2,1080)\, {\rm and}\,  (0,120)$.
In all  cases below we shall take the finest partitions in which a magnetization belongs into a well defined value from $M=-16,\dots M=16$.  Usually the interval is less than 17 and currently the rank of the matrices ranges from $K=2$ up to $K=9$.

The eigenvalues and the invariant probability distributions (the corresponding Eigenvectors associated to the unique unitary Eigenvalue) of these matrices are:

\begin{table*}[h!]
\centering
 \begin{tabular}{|c||c|c|c|c|c| }
  \hline  
   \hline
    $x_{t=0}$ & $\left(
\begin{array}{cccc}
 1 & 1 & 1 & 1 \\
 1 & 1 & 1 & 1 \\
 1 & 1 & 1 & 1 \\
 1 & 1 & 1 & 1 \\
\end{array}
\right)$   &$\left(
\begin{array}{cccc}
 1 & 1 & 1 & 1 \\
 1 & 1 & 1 & 1 \\
 1 & 1 & 1 & 1 \\
 1 & -1 & 1 & -1 \\
\end{array}
\right)$ & $\left(
\begin{array}{cccc}
 1 & 1 & 1 & 1 \\
 1 & 1 & 1 & 1 \\
 1 & 1 & 1 & 1 \\
 1 & -1 & -1 & -1 \\
\end{array}
\right)$ & $\left(
\begin{array}{cccc}
 1 & 1 & 1 & 1 \\
 1 & 1 & 1 & 1 \\
 1 & 1 & 1 & -1 \\
 -1 & -1 & -1 & -1 \\
\end{array}
\right) $ & $\left(
\begin{array}{cccc}
 1 & 1 & 1 & 1 \\
 1 & 1 & 1 & 1 \\
 1 & 1 & 1 & -1 \\
 -1 & 1 & -1 & -1 \\
\end{array}
\right)$    \\  \hline
     $y_{t=0}$ & $\left(
\begin{array}{cccc}
 1 & 1 & 1 & 1 \\
 1 & 1 & 1 & 1 \\
 1 & 1 & 1 & 1 \\
 1 & -1 & 1 & -1 \\
\end{array}
\right)$   &$ \left(
\begin{array}{cccc}
 1 & 1 & 1 & 1 \\
 1 & 1 & -1 & 1 \\
 1 & 1 & 1 & -1 \\
 1 & 1 & -1 & 1 \\
\end{array}
\right) $ &$\left(
\begin{array}{cccc}
 1 & 1 & 1 & 1 \\
 1 & 1 & -1 & 1 \\
 -1 & 1 & 1 & -1 \\
 1 & 1 & -1 & -1 \\
\end{array}
\right) $ & $\left(
\begin{array}{cccc}
 1 & 1 & 1 & 1 \\
 1 & -1 & -1 & -1 \\
 1 & 1 & -1 & -1 \\
 1 & 1 & -1 & -1 \\
\end{array}
\right) $ &    $\left(
\begin{array}{cccc}
 1 & 1 & 1 & 1 \\
 1 & -1 & -1 & -1 \\
 1 & -1 & -1 & -1 \\
 1 & 1 & -1 & -1 \\
\end{array}
\right)$    \\  \hline
   $E$ & $-24$   &$-18$ &$-8 $ & -2 & $0$    \\  \hline  
    $E/N$ & $-1.5$   &$-1.125$ &$-0.5 $ & -0.125& $0$    \\  \hline  
  $ T$ & 6  & $  54 $  & $ 270 $ & 1080 &120  \\ \hline 
   $ I(M)$ & $[12,16]$ &$[8,12]$ & $[2,10]$&$[2,6]$&$[-8,8]$  \\ \hline 
   $K$ & $2$ &$3$ & $5$&$5$&$4$  \\ \hline 
  $ \hat W $&$\left(
\begin{array}{cc}
 \frac{1}{2} & 1 \\
 \frac{1}{2} & 0 \\
\end{array}
\right) $ & $\left(
\begin{array}{ccc}
 0 & \frac{1}{6} & 0 \\
 1 & \frac{1}{2} & 1 \\
 0 & \frac{1}{3} & 0 \\
\end{array}
\right) $ & $\left(
\begin{array}{ccccc}
 0 & \frac{1}{14} & \frac{1}{17} & 0 & 0 \\
 \frac{1}{2} & \frac{5}{14} & \frac{5}{17} & \frac{3}{10} & 0 \\
 \frac{1}{2} & \frac{5}{14} & \frac{6}{17} & \frac{3}{10} & 1 \\
 0 & \frac{3}{14} & \frac{3}{17} & \frac{2}{5} & 0 \\
 0 & 0 & \frac{2}{17} & 0 & 0 \\
\end{array}
\right) $  &$\left(
\begin{array}{ccccc}
 0 & \frac{1}{26} & \frac{1}{24} & \frac{1}{14} & 0 \\
 \frac{1}{4} & \frac{5}{26} & \frac{7}{24} & \frac{9}{28} & \frac{1}{2} \\
 \frac{1}{4} & \frac{7}{26} & \frac{1}{4} & \frac{2}{7} & \frac{1}{4} \\
 \frac{1}{2} & \frac{9}{26} & \frac{1}{3} & \frac{1}{4} & \frac{1}{4} \\
 0 & \frac{2}{13} & \frac{1}{12} & \frac{1}{14} & 0 \\
\end{array}
\right) $  &  $ \left(
\begin{array}{cccc}
 0 & \frac{1}{20} & \frac{1}{14} & 0 \\
 \frac{1}{3} & \frac{3}{10} & \frac{9}{28} & \frac{4}{9} \\
 \frac{2}{3} & \frac{9}{20} & \frac{3}{7} & \frac{5}{9} \\
 0 & \frac{1}{5} & \frac{5}{28} & 0 \\
\end{array}
\right) $   \\        \hline 
    ${\bm f}_{eq} $&$\left( \begin{array}{c} 2/3 \\ 1/3 \end{array}\right) $ & $\left( \begin{array}{c} 1/9\\ 2/3\\ 2/9  \end{array}\right) $ & $\left( \begin{array}{c}   2/45\\ 14/45\\ 17/45\\ 2/9\\ 2/45 \end{array}\right) $  &$\left( \begin{array}{c}  2/45 \\ 13/45 \\ 4/15 \\ 14/45 \\ 4/45  \end{array}\right) $  & $\left( \begin{array}{c} 1/20\\ 1/3\\7/15\\ 3/20 \end{array}\right) $  \\  \hline 
        $\lambda$ &$\left\{ \begin{array}{c} 1\\ -1/2 \end{array}\right\} $  &$\left\{ \begin{array}{c}1\\ -1/2\\ 0 \end{array}\right\} $  & $\left\{  \begin{array}{c}   1\\ 0.314\\ -0.282\\ 0.183\\  -0.104 \end{array}\right\} $  &$\left\{ \begin{array}{c}  1\\ -0.225\\ -0.137\\ 0.073\\ -0.019  \end{array}\right\} $ & $\left\{ \begin{array}{c}  1\\  -1/4\\ -0.0607\\ 0.0392  \end{array}\right\} $   \\   \hline 
       Reversible & Yes & Yes& Yes & Yes &  Yes  \\ \hline 
  \hline
 \end{tabular}
  \caption{\label{Example4x4} Summary of the coarse-graining procedure for Q2R in a $4\times 4$ system and for a given energy $E$ and period $T$.   }
\end{table*}

We notice that the case $E=0$ is showed up in the finest partition. Actually, in the finest partition this case  the probability transfer matrix has a  rank $K=9$. But it has three zero eigenvalues and two complex one. We interpret that the fine coarse graining is not a good partition. This partition must no be an invariant measure as required in Ref.\cite{nicolis}. Why does this happen in the present case?, and it does not happen in other needs to be elucidated.

\end{document}